\documentclass[prstab,prl,preprintnumbers,amsmath,amssymb,runinaddress,preprint]{revtex4}
\usepackage[dvips]{graphicx}
\usepackage{dcolumn}
\usepackage{bm}

\begin{document}

\title{Acoustic Casimir Pressure for Arbitrary Media}

\author{J. B\'{a}rcenas}
\affiliation{Instituto
de F\'{i}sica, Universidad Nacional Aut\'{o}noma de M\'{e}xico, Apartado Postal 20-364 \\
Ciudad Universitaria, M{\'e}xico, 01000, M\'{e}xico.}

\author{L. Reyes}
\affiliation{Instituto
de F\'{i}sica, Universidad Nacional Aut\'{o}noma de M\'{e}xico, Apartado Postal 20-364 \\
Ciudad Universitaria, M{\'e}xico, 01000, M\'{e}xico.}
  \author{R. Esquivel-Sirvent}
  \affiliation{Instituto
de F\'{i}sica, Universidad Nacional Aut\'{o}noma de M\'{e}xico, Apartado Postal 20-364 \\
Ciudad Universitaria, M{\'e}xico  01000, M\'{e}xico.}
\maketitle

\begin{center} Abstract\\ \end{center}
In this paper we derive a general expression for the acoustic Casimir pressure  between two  parallel slabs made of arbitrary materials and whose acoustic reflection coefficients are not equal.  The formalism is based on the calculation of the local density of modes using a Green's function approach. 
 The results for the Casimir acoustic pressure are generalized to a sphere/plate configuration using the proximity theorem.\\
pacs: 43.25.Qp, 43.10.Ln

\newpage

\section{Introduction}
 
Due to quantum vacuum fluctuations, two parallel neutral plates will attract each other. This phenomenon is known as the Casimir force \cite{casimir}.  
 Although a small force,  it has been measured accurately using torsion balances, atomic force microscopes and micro mechanical oscillators \cite{lamoreaux, mohideen1,mohideen2,chan}. An acoustic analog to the Casimir effect was reported in 1996 by Larraza and collaborators \cite{laraza,laraza2, laraza3}, where two parallel plates, placed at a distance $L$,   were subjected to a  broadband noise back ground.  The plates were observed both to attract and to repel each other, depending on the separation distance and the broadband noise cutoff frequencies. Following Casimir's method \cite{casimir}, a theory for the acoustic Casimir force was developed assuming perfectly reflective plates with approximations that turned out to be valid for the frequency range, material and plate thickness that were used in the experiment. In this work we derive a general expression for  the Acoustic Casimir pressure for materials with arbitrary impedances by calculating the density of modes between the plates using the Green's function formalism borrowed from the electromagnetic case. We also present an acoustic analog to the proximity theorem to calculate the Casimir pressure between a sphere and a plate.

\section{Theory}

Consider two  different parallel slabs labeled $i=1,2$ of thickness $d_{1,2}$, separated by a distance $L$ along the $z$ axis.  The slabs are parallel to the $x-y$ plane and have an arbitrary acoustic reflectivity $r_{1,2}$ (Fig. 1a). 

For a perfect acoustic reflector, the radiation pressure 
 of a wave  of intensity $I$ and speed $c$ impinging  on the slab is given by \cite{landau,laraza}
 
\begin{equation}\label{Pr}
P= \frac{2 I }{c} \cos^{2}(\theta),
\end{equation}
where $\theta$ is the angle of incidence.  

As in Ref. \cite{laraza,laraza2,laraza3} we consider broadband acoustic noise of constant spectral intensity $I_{\omega}$ in the frequency interval $[\omega_1,\omega_2]$,  and its spectral representation in the wave vector space

\begin{equation}
I_{k}=\frac{cI_{\omega}¥}{4 \pi k^{2}¥} , 
\end{equation}
where $k^2=(\omega/c)^2=k_x^2+k_y^2+k_z^2$. 

The total radiation pressure on a plate for non perfect reflectors, using Eq. (\ref{Pr}) is
\begin{equation}
P_o=\frac{I_{\omega}}{2 \pi^2}\int dk_x dk_y dk_z \frac{k_z^2}{k^4}. 
\label{p0}
\end{equation}

 Between the plates,  the total radiation pressure$P_{in}$ is determined by the allowed modes that satisfy the boundary conditions at the plate surfaces. If the plates are large enough,   $k_x$ and $k_y$ take on continuous values.  For perfect reflectors the normal component of the wave vector takes the values $k_z=n\pi/L$, where $n$ is an integer. Thus, the calculation of the energy density is reduced to integration over $kx$ and $ky$ and summation over $n$.  The Casimir pressure results from the difference $P_{in}-P_o$.
	
	For arbitrary materials the mode summation is no longer direct, since we are no longer allowed to specify Dirichlet boundary conditions that restrict the allowed modes, and it becomes necessary to calculate the total density of modes ${\mathcal D}(k_z)$ and integrate over all wave vector space.   To do this we use a Green's function approach. 
 
    The wave equation for the velocity potential can be written as the eigenvalue equation $-\partial^2_z \phi=k_z^2 \phi$, with eigenvalues $k_z^2$.  Let $\lambda_n$ be an eigenvalue for the eigenfunction $\phi_n$.  
     In terms of the velocity potential $\phi$, the particle velocity $v_z$ and the fluid density $\rho$ and with the definitions $v_z=\partial_z \phi_n$ and $p=-\rho \partial_t \phi_n$ we can write the normal component of the wave stress tensor \cite{lee2} as 
    \begin{equation}\label{wf}
    w_n=\frac{\rho}{2}\left ( (\partial_z \phi_n)^2 + k_z^2 \phi_n^2 \right ).
    \end{equation}
    
    The total contribution to the stress tensor is obtained by adding up all modes and integrating over all possible values of $k_z^2$. 
        \begin{equation}\label{wmodes}
    w=\frac{\rho}{2}\int dk_z^2 \sum_n \delta(k_z^2-\lambda_n)\left ( (\partial_z \phi_n)^2 + k_z^2 \phi_n^2 \right )    
    \end{equation}
where  we have assumed an harmonic behavior of the potential $\phi$ and the Dirac's delta function is introduced since only the eigenmodes contribute to $w$.  Now, using the identity 
     \begin{equation}\label{deltas}
     \frac{1}{k_z^{+2} -\lambda_n}= P\frac{1}{k^2_z-\lambda_n}-i \pi \delta(k^2_z-\lambda_n),
     \end{equation}
     with $k_z^{+2}=\lim_{\eta \rightarrow 0}(k^2+i \eta)$ we can write Eq.(\ref{wmodes}) as
       \begin{equation}\label{wd2}
    w=\frac{\rho}{2}\int dk_z^2 (-\frac{1}{\pi})Im \sum_n  \frac{1}{k_z^{+2}-\lambda_n}\left ( (\partial_z \phi_n)^2 + k_z^2 \phi_n^2 \right ). 
     \end{equation}
 In this equation, we can identify the spectral representation (or eigenfunction expansion) of the Green's function and its derivative \cite{mrs}, and we interpret the quantity
  \begin{equation}\label{d3}
      \mathcal {D}_{k^2_z}=-\frac{1}{\pi}Im (G(z,z)+\partial _z \partial_z G(z,z)), 
     \end{equation}
as the density of modes.  Another way of understanding the result is as follows. The basic definition of denisty of modes in terms of the Green's funtion is obtained from  Eq. (\ref{deltas}) as $-Im G(z,z)/\pi$.  The acoustic pressure obeys the wave equation and with th eappropriate boundary conditions, we can obtain the Green's function $G_p$ and thus the density of modes. From the acoustic stress tensor component (Eq.(\ref{wf})), besides the pressure field there is a contribution from the velocity field. Let this field have an associated Green's function $G_v$ . The total density of modes of the system  $-Im(G_p+G_v)/\pi$ of both fields.  Writing the pressure and velocity in terms of the scalar potential yields Eq. (\ref{d3}). This is equivalent to what happens in zero point Casimir effect where the density of modes comes from adding the contribution of the electric field plus that due to the magnetic fields, and both fields are related through a constitutive equation (Maxwell's equations). 

To construct the Green's function 
  for the velocity potential we can use the standard definition 
   \begin{equation}\label{green}
G _{k^2}(z,z')= \frac{\phi^{<}(z_<) \phi^{>}(z_{>})}{W},
\end{equation}
where $W$ is the Wronskian and  
\begin{equation}\label{ondas}
\begin{split}
& \phi^{<}(z)= e^{-ik_zz}+r_1e^{ik_zz}\\
&  \phi^{>}(z)= e^{ik_z(z-L)}+r_2e^{-ik_z(z-L)},
\end{split}
\end{equation}
are the solutions to the one dimenional wave equation where the super-index ($<,>$)represents the smaller and larger of $z$ and $z'$ respectively. 
  
Substitution of the potentials (Eq.(\ref{ondas})) into Eq.(\ref{green})  and Eq. (\ref{d3}) yields
the local  density of modes
\begin{equation}\label{distra}
  \mathcal {D}_{k^2_z}=  \frac{1}{2 k_z \pi} Re\left[ \frac{1+r_1 r_2 e^{2ik_zL}}{1-r_1 r_2e^{2 i k_z L}}\right ],
\end{equation}
where we have obviated the dependence of the reflectivities with wave vector.  
  The density of states Eq.(\ref{distra}) was obtained from the Green's function for the Helmholtz equation with eigenvalues $k_z^2$. However, we are interested in the density of states for $k_z$.  This is simply obtained from $  \mathcal {D}_{k^2_z}=d(k^2_z)   \mathcal {D}_{k_z} =2 k_z   \mathcal {D}_{k^2_z}$, or
   \begin{equation}\label{distraf}
  \mathcal {D}_{k_z}=  \frac{1}{ \pi} Re\left[ \frac{1+r_1 r_2 e^{2ik_zL}}{1-r_1 r_2 e^{2 i k_z L}}\right ],
\end{equation}
  
 The radiation pressure due to the inside modes can now be written as
\begin{equation}
P_{in}=\frac{I_{\omega}}{4 \pi}\int dk_x dk_y dk_z {\mathcal D}_{k_z}\frac{k_z^2}{k^4}.
\label{pin}
\end{equation}
  In the limit of perfect reflectors $r\rightarrow 1$ the density of states becomes $ \frac{k_0}{\pi} \delta(k_z-n k_0)$ where $k_0=\pi/L$.  Thus, from Eq. (\ref{pin}) the pressure due to all the modes is
\begin{equation}
P_{in}=\frac{k_0 I_{\omega}}{2 \pi^2}\int dk_x dk_y dk_z \sum_n \delta (k_z-n k_0) \frac{k_z^2}{k^4},
\end{equation}
or 
\begin{equation}
P_{in}=\frac{k_0^3 I_{\omega}}{2 \pi}\int dk_x dk_y  \sum_n\frac{n^2}{(k_x^2+k_y^2+n^2 k_0^2)^2},
\end{equation}
that is the same as that obtained by Larraza and collaborators \cite{laraza,laraza2,laraza3}.
 Finally, the acoustic Casimir force per unit area   
$f=P_{in}-P_{out}$ takes the form
\begin{equation}
f=\frac{I_{\omega}}{2 \pi^2}Re \left(\int dk_x dk_y dk_z \frac{k_z^2}{k^4}(\frac{1}{\xi-1}) \right ), 
\label{deltap}
\end{equation}
where $\xi=(r_1 r_2 \exp{(2 i k_z L)})^{-1}$. Notice that it is enough to know the separation between the slabs and the reflectivities to determine the acoustic Casimir force. 

  To illustrate the application of Eq. (\ref{deltap}), in Fig. 2 we plot the force {\it vs} separation  for two identical slabs with constant reflectivities $r=1,.8,.7$.  In all cases the force  goes from attractive to repulsive as the separation increases. The magnitude of the force is not only related to the reflectivity but also to the finite bandwidth being used. If the band width extends from zero to infinity, the acoustic Casimir pressure for a perfect reflector $ -\pi I_{\omega}/4 L$ is always attractive. From Eq. (\ref{deltap}) if we integrate over all frequencies the force is also  always attractive and as the reflectivity decreases the force does too \cite{footnote1} in all cases.  Without loss of generality we have assumed a constant value of $r$ within the bandwidth under consideration.  However, the formalism is valid even when the reflectivity shows a strong dependence with frequency.  The bandwidth and intensity used in these calculations are the same as in Laraza \cite{laraza}. 
   Interestingly, even if we consider a finite frquency bandwidth it is possible to obtain a purely attractive force if we consider the force between a surface with reflectivity $r_1=1$ and a pressure release surface $r_2=-1$. In this limit the force is constant and  always attractive since $D_{k_z}\rightarrow0$, as can be seen from Eq. (\ref{distraf}) and the external  pressure field  pushes the plates together.

\section{The proximity theorem in acoustics}

Practical measurements of the Casimir force due to zero point energy fluctuations are done between a large sphere and a plane due to the difficulty of keeping two plates parallel at the submicron scale \cite{lamoreaux, mohideen1,mohideen2,chan}. 
The force between a large sphere and a plane (see Fig. 1b) is calculated using the proximity theorem\cite{proximity} or Derjaguin approximation\cite{derja}, which states that 
\begin{verse}
 {\it  the force between two smooth surfaces as a function of the separation degree of freedom is proportional to the interaction potential per unit area $\mathcal{E}$ between two flat surfaces, the proportionality factor being  $2 \pi$ times the reciprocal  of the square root  of the Gasussian cuadrature  of the gap width function  at the point of closest approach .}
\end{verse}
For a sphere-plane system, the force  $F_{sp}$ is obtained from the Casimir free energy per unit area between two parallel plates $\cal{E}$ as 
\begin{equation}\label{pt}
F_{sp}= 2 \pi R \cal{E}, 
\end{equation}
where $R$ is the radius of the sphere.  The proximity theorem is valid  
provided $L/R<1$, being $L$ the closest distance between the surface, although the limit of $L\rightarrow 0$ can be described by the proximity theorem. A current problem \cite{ceci}, is that there are no bounds on how big $L/R$ has to be in order to obtain the correct result.  Experimentally this becomes difficult at the submicron scale. The acoustic analog of the Casimir force provides a simpler (not necessarily easier) way of solving this problem, since as shown by Laraza \cite{laraza}
the scale of the acoustic experiments allows a more precise control of the involved parameters.  Furthermore, the proximity theorem is valid for any interaction.  In the acoustic case,  the free energy per unit area for parallel plates is 

\begin{equation}
\label{fren}
\mathcal{E}=\frac{I_{\omega}}{4 \pi^2}\int dk_x dk_y dk_z \frac{k_z}{k^4}Re\left (ln(1-r_1r_2e^{2 i k_z L}))\right ). 
 \end{equation}
 This expression for $\mathcal{E}$ is such that the force (Eq. (\ref{deltap})) is given by $f=-\partial \mathcal{E}/\partial L$.  Thus, the force between a sphere and a plane is 

\begin{equation}
\label{fps}
f_{ps}=\frac{R I_{\omega}}{ \pi}\int dk_x dk_y dk_z \frac{k_z}{k^4}Im\left (ln(1-r_1r_2e^{2 i k_z L}))\right ).
 \end{equation}
    To show the application of this approximation, in Fig. 3 we have plotted the force between two parallel plates and the force between a $20 cm$ sphere and a plate.  We observe that although the  proximity theorem  gives a correct behavior and overall oreder of magnitude for the force,  the region in the limit of $L$ approaching zero is not well described.  

\section{Conclusions}

We have derived a general expression for the Acoustic  Casimir force  between two parallel slabs 
with arbitrary acoustic  properties characterized by the reflection 
coefficients of the material.  We also extended our results to include the force between a sphere and a plane. 
 The expression we obtain for the Casimir force is convenient for 
calculations since it depends mainly on the reflection coefficients that can be obtained 
straightforwardly in theoretical computations or through experimental work. 
 Our approach is analogous to the electromagnetic dielectric case, so this formalism is equivalent to the Liftshitz formula \cite{raul,nos4}. 
 In the limit of a  perfectly reflective plate our results agree with those of Laraza \cite{laraza}.   This formalism can be extended to the case of highly porous materials or viscous propagation media, although the calculations involved can be of increasing difficulty. It must be pointed out that for the case when the material is deformed by the wave, this density of states approach is no longer valid: the reflection coefficient is angular dependent, and since the angle itself is time dependent the use of a static density of states would be incorrect. Also, we have excluded the possible effects of roughness.   

The crucial difference between the system we consider and the original treatment \cite{laraza} is the inclusion of the density of states through the Green's function method. The analytical interpretation of the density function gives a deeper insight into what really happens in a non perfect reflector. 
For a perfect reflector the density of modes consists of a series of Dirac's deltas.  As the reflectivity decreases from unity,  the resonance bands increase in width which is heuristically equivalent to a spatial diffusion of the nodes that appear inside the resonant cavities.

Although the use of perfectly reflectiving plates is a good approximation in some experimental situations,  this is not the case  for other bodies (such as rubber, as an extreme example) hence our efforts to broaden the horizon of application.  As an example, we have consider the possibility of using acoustic experiments to prove the validity of the proximity theorem. Additionally, this treatment could allow for a larger range of experimental vs. theoretical comparison in this field where, as noted by Larraza et al, the possibility of direct technological application of the acoustic Casimir effect is considerable.

acknowledgements: Partial support provided by DGAPA-UNAM project IN116002-2.

 \newpage
 
 \begin{figure}[h]
\caption{a)  Geometry and coordinate system for the two parallel plate configuration and b) for the sphere-plane configuration.}
 \label{fig1}
\end{figure}

\begin{figure}[h]
\caption{Acoustic Casimir force between two parallel plates for different values of the reflectivity $r$ assuming both plates are equal. The values of $r$ are indicated in the figure. The intensity and bandwidth are the same as in the experiments of Laraza  \cite{laraza}).}
 \label{fig2}
\end{figure}

\begin{figure}[h]
\caption{Acoustic Casimir force between a sphere (R=0.2 m) and a plane for reflectivity $r=1$ (dashed line). For comparison the force between two parallel plates is also shown (solid line).}
 \label{fig3}
\end{figure}


\begin{thebibliography}{99}
\bibitem{footnote1}  When the bandwidth extends from zero to infinity it is possible to evaluate the force by mapping the real frequency axis $\omega$ to $i \omega$, making the  integral numerically stable.  See for example Ref. [\cite{nos4}].
\end{thebibliography}
 \end{document}